\documentclass[journal]{IEEEtran}
\usepackage{mathrsfs}
\usepackage{cite}
\usepackage{amsmath}
\usepackage{amssymb}
\usepackage{stfloats}
\usepackage{graphicx}
\usepackage{setspace}
\usepackage{xcolor}

\ifCLASSOPTIONcompsoc
\usepackage[tight,normalsize,sf,SF]{subfigure}
\else
\usepackage[tight,footnotesize]{subfigure}
\fi

\begin{document}
\title{Spatial Covariance Matrix Reconstruction for DOA Estimation in Hybrid Massive MIMO Systems with Multiple Radio Frequency Chains}
\author{Yinsheng~Liu, Yiwei~Yan, Li~You, Wenji~Wang, and Hongtao~Duan.

\thanks{Yinsheng Liu is with State Key Laboratory of Rail Traffic Control and Safety, Beijing Jiaotong University, Beijing 100044, China, and National Mobile Communications Research Laboratory, Southeast University, Nanjing 210096, China (e-mail: ys.liu@bjtu.edu.cn).}
\thanks{Yiwei Yan is with School of Electronic and Information Engineering, Beijing Jiaotong University, Beijing 100044, China, (e-mail:19120159@bjtu.edu.cn).}
\thanks{Li You and Wenjin Wang are with National Mobile Communications Research Laboratory, Southeast University, Nanjing 210096, China, and Purple Mountain Laboratories, Nanjing 211100, China (e-mail:liyou@seu.edu.cn,wangwj@seu.edu.cn).}
\thanks{Hongtao Duan is with Beijing radio monitoring station of State Radio Monitoring Center (SRMC), Beijing 100037, China (e-mail: duanht@srrc.org.cn).}

\thanks{Corresponding author: Yinsheng Liu.}
}
\maketitle
\begin{abstract}
Multiple signal classification (MUSIC) has been widely applied in multiple-input multiple-output (MIMO) receivers for direction-of-arrival (DOA) estimation. To reduce the cost of radio frequency (RF) chains operating at millimeter-wave bands, hybrid analog-digital structure has been adopted in massive MIMO transceivers. In this situation, the received signals at the antennas are unavailable to the digital receiver, and as a consequence, the spatial covariance matrix (SCM), which is essential in MUSIC algorithm, cannot be obtained using traditional sample average approach. Based on our previous work, we propose a novel algorithm for SCM reconstruction in hybrid massive MIMO systems with multiple RF chains. By switching the analog beamformers to a group of predetermined DOAs, SCM can be reconstructed through the solutions of a set of linear equations. In addition, based on insightful analysis on that linear equations, a low-complexity algorithm, as well as a careful selection of the predetermined DOAs, will be also presented in this paper. Simulation results show that the proposed algorithms can reconstruct the SCM accurately so that MUSIC algorithm can be well used for DOA estimation in hybrid massive MIMO systems with multiple RF chains.
\end{abstract}


\section{Introduction}
Direction-of-arrival (DOA) estimation has been widely used in wireless communications because it can determine the directions of unknown signal sources \cite{TETuncer,FShu1}. As an important DOA estimation approach, multiple signal classification (MUSIC) has gained a lot of attention due to its super-resolution property in the presence of multiple signals \cite{ROSchmidt2}.\par

Massive multiple-input multiple-output (MIMO) is one of the most important enabling technologies in 5G and Beyond 5G communication systems \cite{EGLarsson}. Due to a large number of antennas, massive MIMO is essential to millimeter-wave bands because the large array gain can compensate for the high path loss. With the help of massive MIMO, the frequency resources at millimeter-wave bands can be exploited efficiently in 5G and Beyond 5G communication systems \cite{WRoh,LYou}.\par

To reduce the number of radio frequency (RF) chains, hybrid structure has been adopted for massive MIMO operating at millimeter-wave bands \cite{OEAyach1,VVen,CLin}. In hybrid systems, one RF chain is connected to multiple antennas, so that the number of RF chains can be greatly reduced. However, in hybrid massive MIMO, the received signals are first fed to the analog phase shifters and then combined in the analog domain before sent to the digital receiver. Consequently, the received signals at the antennas are unavailable to the digital receiver, and the spatial covariance matrix (SCM), which is essential in MUSIC algorithm, cannot be obtained using the traditional sample average approach \cite{SChuang}. As MUSIC algorithm is not applicable in hybrid systems, a straightforward strategy for DOA estimation is to search for the direction with the maximum received power \cite{FShu,KAgha,DHu}, which, however, is restricted by Rayleigh limitation \cite{ROSchmidt2}. Rayleigh limitation refers to the limitation on the angle resolution in the presence of multiple signals. It is proportional to the aperture and thus a large number of antennas are required for better resolution.\par

To make use of the super-resolution property of MUSIC algorithm, we have developed a beam sweeping algorithm for SCM reconstruction in massive MIMO systems with single RF chain \cite{SLi}. In this paper, the beam sweeping algorithm is improved to enable SCM reconstruction in massive MIMO systems with multiple RF chains. First, the overall SCM is divided into a number of sub-SCMs. Then, similar to \cite{SLi}, by switching the directions of multiple beamformers to predetermined DOAs in turn, each sub-SCM can be reconstructed through solving a set of linear equations.
Furthermore, it shows that there are a lot of redundant calculation in the basic beam sweeping algorithm, because many repeated entries exist in the SCM. Based on this observation, the algorithm optimization is further investigated in this paper. First, a low-complexity beam sweeping algorithm is presented, where the computation complexity can be reduced to be linearly proportional to the number of antennas. Then, the selection of predetermined DOAs will be also optimized in this paper. Using the optimized selection, the number of required predetermined DOAs can be reduced significantly, so that the SCM reconstruction procedure can be accomplished in a shorter interval. Simulation results have also been presented to demonstrate the proposed algorithms.

The rest of this paper is organized as follows. In Section II, signal model for hybrid massive MIMO is introduced. In Section III, beam sweeping algorithm is presented and the algorithm optimization is shown in Section IV. Simulation results can be found in Section V and the conclusions are drawn in Section VI.

\section{System Model}
\subsection{Signal Model}
\begin{figure}
  \centering
  \includegraphics[width=3.5in]{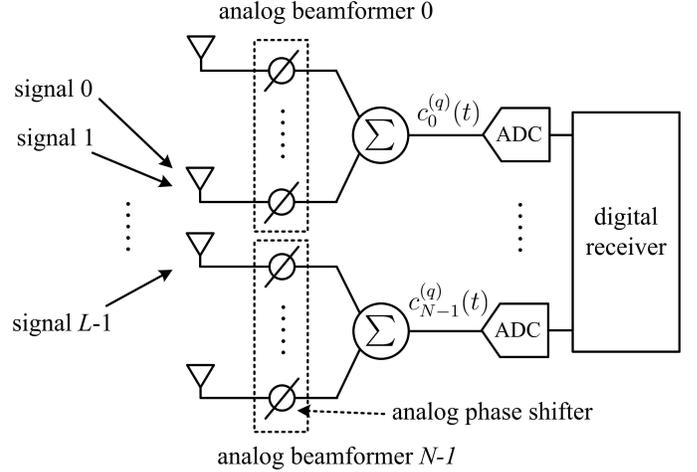}\\
  \caption{For a hybrid massive MIMO with multiple RF chains, each RF chain is connected to a subset of the antennas.}\label{system}
\end{figure}

As in Fig.~\ref{system}, consider a hybrid massive MIMO system composed of a uniform linear array (ULA) with $M$ antennas and $N$ RF chains. Denote $y_{m,n}(t)$ to be the received signal at the $m$-th antenna of the $n$-th RF chain. Since each RF chain is connected to $\frac{M}{N}$ antennas, we have $m = 0,1,\cdots,\frac{M}{N} - 1$ and $n = 0,1,\cdots,N - 1$. Then, the received signal vector by the $n$-th RF chain $\boldsymbol{y}_n(t)=[y_{0,n}(t),y_{1,n}(t),\cdots,y_{\frac{M}{N}-1,n}(t)]^{\mathrm{T}}$ can be represented as
\begin{align}\label{signalmodel}
\boldsymbol{y}_n(t)=\sum_{l=0}^{L-1}\boldsymbol{a}_n(\theta_l)x_l(t)+\boldsymbol{z}_n(t),
\end{align}
where $x_l(t)$'s ($l=0,1,\cdots,L - 1)$ are $L$ narrow-band signals impinging from far field onto the array, $\theta_l$ is the DOA of $x_l(t)$, $\boldsymbol{z}_n(t)$ denotes the additive Gaussian noise vector with $\mathrm{E}\{\boldsymbol{z}_n(t)\boldsymbol{z}_n^{\mathrm{H}}(t)\}=N_0\boldsymbol{I}_{\frac{M}{N}}$ where $N_0$ is the noise power and $\boldsymbol{I}_{\frac{M}{N}}$ is an $\frac{M}{N}\times \frac{M}{N}$ identity matrix, and $\boldsymbol{a}_n(\theta_l)$ is the $\frac{M}{N}\times 1$ steering vector corresponding to the $n$-th RF chain with the $m$-th entry given by
\begin{align}
a_{m,n}(\theta_l)=e^{j2\pi\cdot\frac{d}{\lambda}\cdot\sin\theta_l\cdot (n\frac{M}{N} + m)},
\end{align}
where $d=\frac{\lambda}{2}$ denotes the antenna distance and $\lambda$ is the wave length. Take all the RF chains into account, then the overall received signal vector, the overall steering vector, and the overall additive noise vector can be represented as
\begin{align}\label{overall}
\boldsymbol{y}(t)&=[\boldsymbol{y}_{0}^{\mathrm{T}}(t),\boldsymbol{y}_{1}^{\mathrm{T}}(t),\cdots,\boldsymbol{y}_{N-1}^{\mathrm{T}}(t)]^{\mathrm{T}},\\
\boldsymbol{a}(\theta_l)&=[\boldsymbol{a}_{0}^{\mathrm{T}}(\theta_l),\boldsymbol{a}_{1}^{\mathrm{T}}(\theta_l),\cdots,\boldsymbol{a}_{N-1}^{\mathrm{T}}(\theta_l)]^{\mathrm{T}},\\
\boldsymbol{z}(t)&=[\boldsymbol{z}_0^{\mathrm{T}}(t),\boldsymbol{z}_1^{\mathrm{T}}(t),\cdots,\boldsymbol{z}_{N-1}^{\mathrm{T}}(t)]^{\mathrm{T}},
\end{align}
respectively. Accordingly, we have
\begin{align}\label{signalmodel}
\boldsymbol{y}(t)=\sum_{l=0}^{L-1}\boldsymbol{a}(\theta_l)x_l(t)+\boldsymbol{z}(t).
\end{align}

Denote $\boldsymbol{R}=\mathrm{E}\{\boldsymbol{y}(t)\boldsymbol{y}^{\mathrm{H}}(t)\}$ to be the overall SCM, then using (\ref{overall}), the overall SCM can be divided into
\begin{align}\label{divide}
\boldsymbol{R}=\left[\begin{array}{ccc}
                       \boldsymbol{R}_{0,0} & \cdots & \boldsymbol{R}_{0,N-1} \\
                       \vdots & \ddots & \vdots \\
                       \boldsymbol{R}_{N-1,0} & \cdots & \boldsymbol{R}_{N-1,N-1}
                     \end{array}
\right],
\end{align}
where $\boldsymbol{R}_{n_1,n_2}=\mathrm{E}\{\boldsymbol{y}_{n_1}(t)\boldsymbol{y}_{n_2}^{\mathrm{H}}(t)\}$ is the $(n_1,n_2)$-th sub-SCM. Assuming $L$ signals are mutually independent with zero means and the power of the $l$-th signal is $\mathrm{E}\{|x_l(t)|^2\}=\sigma_l^2$, then the $(n_1,n_2)$-th sub-SCM and the overall SCM will be
\begin{align}
\boldsymbol{R}_{n_1,n_2}&=\sum_{l=0}^{L-1}\sigma_l^2\cdot\boldsymbol{a}_{n_1}(\theta_l)\boldsymbol{a}_{n_2}^{\mathrm{H}}(\theta_l) + \delta[n_1-n_2]N_0\boldsymbol{I}_{\frac{M}{N}},\\
\boldsymbol{R}&=\sum_{l=0}^{L-1}\sigma_l^2\cdot\boldsymbol{a}(\theta_l)\boldsymbol{a}^{\mathrm{H}}(\theta_l) +N_0\boldsymbol{I}_M,\label{R}
\end{align}
respectively, where $\delta[\cdot]$ indicates Kronecker Delta function.

\subsection{Review of MUSIC Algorithm}
Denote $\boldsymbol{y}[k]=\boldsymbol{y}(kT_s)$ to be the sample of the received signal where $T_s$ denotes the sampling period. In MUSIC algorithm, $\boldsymbol{y}[k]$'s are assumed to be available in the receiver. In this case, the overall SCM in (\ref{R}) can be estimated using the sample average approach, that is \cite{ROSchmidt2,GMan}
\begin{align}\label{approx}
\boldsymbol{R}\approx \frac{1}{K}\sum_{k=0}^{K-1}\boldsymbol{y}[k]\boldsymbol{y}^{\mathrm{H}}[k],
\end{align}
where $K$ denotes the number of samples. The eigenvalue decomposition of the overall SCM can be given as
$
\boldsymbol{R}=(\boldsymbol{U}_s,\boldsymbol{U}_n)\boldsymbol{\Lambda}_s(\boldsymbol{U}_s,\boldsymbol{U}_n)^{\mathrm{H}},
$
where $\boldsymbol{U}_s$ and $\boldsymbol{U}_n$ denote the orthogonal base vectors corresponding to the signal and the noise subspaces respectively, and $\boldsymbol{\Lambda}_s$ is a diagonal matrix composed of the eigenvalues. Then, unknown DOAs can be determined by searching the peak values of $p(\theta)$,
\begin{align}
p(\theta) = {\|\boldsymbol{U}_n^{\mathrm{H}}\boldsymbol{a}(\theta)\|_2^{-2}},~~~\theta\in[-90^{\mathrm{o}},90^{\mathrm{o}}].\label{Un}
\end{align}
\par

For sample average in (\ref{approx}), $\boldsymbol{y}[k]$ is required to estimate SCM. In this case, the received signals at all antennas should be sent via RF chains to the digital receiver. In hybrid massive MIMO, however, Fig.~\ref{system} shows that only the combination of the entries of $\boldsymbol{y}_n[k]$ can be sent to the digital receiver because the number of RF chains is smaller than the number of antennas. As a consequence, the sample average approach in (\ref{approx}) cannot be used in hybrid massive MIMO systems. In \cite{SLi}, we have developed an algorithm for SCM reconstruction in massive MIMO with single RF chain. In the presence of multiple RF chains, however, each RF chain is connected to only a subset of the antennas. Therefore, the overall SCM cannot be obtained directly using the algorithm in \cite{SLi}.

\section{Beam Sweeping Algorithm}
To reconstruct the overall SCM in the presence of multiple RF chains, we can first reconstruct the sub-SCMs individually, and then the overall SCM can be obtained following (\ref{divide}).\par

To reconstruct $\boldsymbol{R}_{n_1,n_2}$, define $\{\theta^{(0)},\theta^{(1)},\cdots,\theta^{(Q-1)}\}$ as a set of predetermined DOA angles. The analog beamformers switch the beam directions to the predetermined DOAs in turn. For the $q$-th sweeping beam, the predetermined DOA is $\theta^{(q)}$, and thus the steering vector corresponding to the $n$-th RF chain is $\boldsymbol{a}_n(\theta^{(q)})$. The combination of the received signals on the $n$-th RF chain can be represented by
\begin{align}
c_{n}^{(q)}(t)=\boldsymbol{a}_n^{\mathrm{H}}(\theta^{(q)})\boldsymbol{y}_n(t).
\end{align}
From Fig.~\ref{system}, the $n$-th signal combination is sampled before sent to the receiver, and thus the samples of the signal combination can be given by
\begin{align}
c_{n}^{(q)}[k]= c_{n}^{(q)}(kT_s)=\boldsymbol{a}_n^{\mathrm{H}}(\theta^{(q)})\boldsymbol{y}_n[k].
\end{align}\par
Denote $P_{n_1,n_2}^{(q)}$ to be the correlation between the outputs of the $n_1$-th and the $n_2$-th RF chains, that is
\begin{align}\label{corr}
&P_{n_1,n_2}^{(q)}=\frac{1}{K}\sum_{k=0}^{K-1}c_{n_1}^{(q)}[k]c_{n_2}^{(q)*}[k]\nonumber\\
&=\boldsymbol{a}_{n_1}^{\mathrm{H}}(\theta^{(q)})\left(\frac{1}{K}\sum_{k=0}^{K-1}\boldsymbol{y}_{n_1}[k]\boldsymbol{y}_{n_2}^{\mathrm{H}}[k]\right)\boldsymbol{a}_{n_2}(\theta^{(q)}).
\end{align}
If the number of samples is large enough, the sample average in (\ref{corr}) is equivalent to the statistical average, that is
\begin{align}\label{tight}
\boldsymbol{a}_{n_1}^{\mathrm{H}}(\theta^{(q)})\boldsymbol{R}_{n_1,n_2}\boldsymbol{a}_{n_2}(\theta^{(q)})=P_{n_1,n_2}^{(q)}.
\end{align}
Using the $\mathrm{vec}(\cdot)$ operator to (\ref{tight}), the left-hand-side of (\ref{tight}) can be given as
\begin{align}
&\mathrm{vec}\{\boldsymbol{a}_{n_1}^{\mathrm{H}}(\theta^{(q)})\boldsymbol{R}_{n_1,n_2}\boldsymbol{a}_{n_2}(\theta^{(q)})\}\nonumber\\
&=[\boldsymbol{a}_{n_2}^{\mathrm{T}}(\theta^{(q)})\otimes \boldsymbol{a}_{n_1}^{\mathrm{H}}(\theta^{(q)})]\mathrm{vec}(\boldsymbol{R}_{n_1,n_2}),
\end{align}
where we have used the equation (1.10.25) in \cite{XZhangTsingHua}, that is, $\mathrm{vec}(\boldsymbol{BCD})=(\boldsymbol{D}^{\mathrm{T}}\otimes\boldsymbol{B})\mathrm{vec}(\boldsymbol{C})$ with $\otimes$ denoting the Kronecker product.

To proceed, denote $\boldsymbol{r}_{n_1,n_2}=\mathrm{vec}(\boldsymbol{R}_{n_1,n_2})$ and
\begin{align}\label{aKro}
\boldsymbol{a}_{n_1,n_2}^{(q)}=\boldsymbol{a}_{n_2}(\theta^{(q)})\otimes\boldsymbol{a}_{n_1}^*(\theta^{(q)}),
\end{align}
where $(\cdot)^*$ denotes the element-wise conjugation. Apparently, both $\boldsymbol{r}_{n_1,n_2}$ and $\boldsymbol{a}_{n_1,n_2}^{(q)}$ are $\frac{M^2}{N^2}\times 1$ vectors. Then, (\ref{tight}) can be rewritten as
\begin{align}\label{single}
(\boldsymbol{a}_{n_1,n_2}^{(q)})^{\mathrm{T}}\boldsymbol{r}_{n_1,n_2}=P_{n_1,n_2}^{(q)}.
\end{align}
Considering that there are $Q$ predetermined DOA angles, then (\ref{single}) can be extended to a group of linear equations as
\begin{align}\label{equations}
\boldsymbol{A}_{n_1,n_2}\boldsymbol{r}_{n_1,n_2}=\boldsymbol{p}_{n_1,n_2},
\end{align}
where $\boldsymbol{A}_{n_1,n_2}$ is a $Q\times \frac{M^2}{N^2}$ matrix and $\boldsymbol{p}_{n_1,n_2}$ is a $Q\times 1$ vector
\begin{align}
\boldsymbol{A}_{n_1,n_2}=\left[\boldsymbol{a}_{n_1,n_2}^{(0)},\boldsymbol{a}_{n_1,n_2}^{(1)},\cdots,\boldsymbol{a}_{n_1,n_2}^{(Q-1)}\right]^{\mathrm{T}},\\
\boldsymbol{p}_{n_1,n_2}=\left[P_{n_1,n_2}^{(0)},P_{n_1,n_2}^{(1)},\cdots,P_{n_1,n_2}^{(Q-1)}\right]^{\mathrm{T}}.
\end{align}

Then, the $(n_1,n_2)$-th sub-SCM can be reconstructed by solving (\ref{equations}). Similar to \cite{SLi}, a diagonal loading coefficient can be adopted to improve the distribution of the eigenvalues, so that the ill-conditioned result can be avoided. In this case, the vector form of sub-SCM can be obtained as
\begin{align}\label{inverse}
\widehat{\boldsymbol{r}}_{n_1,n_2}=(\boldsymbol{A}_{n_1,n_2}^{\mathrm{H}}\boldsymbol{A}_{n_1,n_2}+\sigma^2\boldsymbol{I}_{\frac{M^2}{N^2}})^{-1}\boldsymbol{A}_{n_1,n_2}^{\mathrm{H}}\boldsymbol{p}_{n_1,n_2},
\end{align}
where $\sigma^2$ denotes the diagonal loading coefficient. Then, the $(n_1,n_2)$-th sub-SCM can be reconstructed through
$
\widehat{\boldsymbol{R}}_{n_1,n_2}=\mathrm{unvec}(\widehat{\boldsymbol{r}}_{n_1,n_2}).
$
Once the sub-SCMs are obtained, the overall SCM can be reconstructed, following (\ref{divide}), as
\begin{align}
\widehat{\boldsymbol{R}}=\left[\begin{array}{ccc}
                       \widehat{\boldsymbol{R}}_{0,0} & \cdots & \widehat{\boldsymbol{R}}_{0,N-1} \\
                       \vdots & \ddots & \vdots \\
                       \widehat{\boldsymbol{R}}_{N-1,0} & \cdots & \widehat{\boldsymbol{R}}_{N-1,N-1}
                     \end{array}
\right].\label{reconstructR}
\end{align}
As in \cite{SLi}, although matrix inversion in (\ref{inverse}) causes a huge computational burden, the operator $(\boldsymbol{A}_{n_1,n_2}^{\mathrm{H}}\boldsymbol{A}_{n_1,n_2}+\sigma^2\boldsymbol{I}_{\frac{M^2}{N^2}})^{-1}\boldsymbol{A}_{n_1,n_2}^{\mathrm{H}}$ can be pre-calculated off-line if the predetermined DOAs are fixed. In this case, matrix inversion can be avoided from on-line calculation, and the computational burden is mainly caused by the matrix-vector production in (\ref{inverse}), which requires $QM^2/N^2$ complex multiplications.
\par
With the reconstructed SCM in (\ref{reconstructR}), unknown DOAs can obtained using (\ref{Un}) if classical MUSIC algorithm is adopted. In addition to the classical MUSIC algorithm, the variants of MUSIC algorithm, such as root-MUSIC \cite{RaoBD}, can be also used. To use root-MUSIC and other variants, we only need to replace sample average approach in those algorithms with the SCM reconstruction algorithm in this paper.

\section{Algorithm Optimization}
Although the beam sweeping algorithm in Section III can reconstruct the SCM successfully, the computation complexity is still huge due to the large dimension of matrix-vector production in (\ref{inverse}). In this section, a low-complexity algorithm will be presented where the dimension of matrix-vector production can be reduced significantly. Based on the low-complexity algorithm, optimized selection of predetermined DOAs is further investigated. Using the optimized predetermined DOAs, the matrix inverse operation can be even avoided.

\subsection{Low-complexity Implementation}
Low-complexity implementation of the beam sweeping algorithm is inspired by the fact that many repeated entries exist in the SCM. Although there are $\frac{M^2}{N^2}$ entries in $\boldsymbol{R}_{n_1,n_2}$, the number of non-repeated unknowns is only $\frac{2M}{N}-1$. Therefore, the computational complexity can be reduced if we only recover the non-repeated unknowns.

Denote the $(m_1,m_2)$-th entry of $\boldsymbol{R}_{n_1,n_2}$, $\boldsymbol{R}_{n_1,n_2}[m_1,m_2]$, to be $\gamma_{n_1,n_2}[m_1-m_2]$. If denote
$\boldsymbol{\gamma}_{n_1,n_2}$ to be a $(\frac{2M}{N}-1)\times 1$ column vector containing all non-repeated unknowns in $\boldsymbol{r}_{n_1,n_2}$ or $\boldsymbol{R}_{n_1,n_2}$, then $\boldsymbol{\gamma}_{n_1,n_2}$ can be given by
\begin{align}
\boldsymbol{\gamma}_{n_1,n_2}=\left(\gamma_{n_1,n_2}\left[1-\frac{M}{N}\right],\cdots,\gamma_{n_1,n_2}\left[\frac{M}{N}-1\right]\right)^{\mathrm{T}}.
\end{align}
Accordingly, $\boldsymbol{r}_{n_1,n_2}$ can be expressed by $\boldsymbol{\gamma}_{n_1,n_2}$ using
\begin{align}\label{compress}
\boldsymbol{r}_{n_1,n_2}=\boldsymbol{E}\cdot\boldsymbol{\gamma}_{n_1,n_2},
\end{align}
where $\boldsymbol{E}$ is an $\frac{M^2}{N^2}\times \left(\frac{2M}{N}-1\right)$ matrix
\begin{align}
\boldsymbol{E}=\left[\begin{array}{ccc}
                                               \boldsymbol{O}_{\frac{M}{N} - 1} & \boldsymbol{I}_{\frac{M}{N}} & \boldsymbol{O}_0 \\
                                              \boldsymbol{O}_{\frac{M}{N} - 2} & \boldsymbol{I}_{\frac{M}{N}} & \boldsymbol{O}_1 \\
                                               \vdots & \vdots & \vdots \\
                                             \boldsymbol{O}_{0} & \boldsymbol{I}_{\frac{M}{N}} & \boldsymbol{O}_{\frac{M}{N} - 1}
                                             \end{array}
\right]\label{matE}
\end{align}
with $\boldsymbol{O}_{m}$ denoting an $\frac{M}{N}\times m$ all-zero matrix. By substituting (\ref{compress}) into (\ref{single}), (\ref{single}) can be rewritten as
\begin{align}\label{bsingle11}
{(\boldsymbol{b}_{n_1,n_2}^{(q)})^{\mathrm{T}}}\boldsymbol{\gamma}_{n_1,n_2}=P_{n_1,n_2}^{(q)},
\end{align}
where $(\boldsymbol{b}_{n_1,n_2}^{(q)})^{\mathrm{T}}$ is a $1\times \left(\frac{2M}{N} - 1\right)$ row vector given by
\begin{align}\label{defb}
(\boldsymbol{b}_{n_1,n_2}^{(q)})^{\mathrm{T}}=(\boldsymbol{a}_{n_1,n_2}^{(q)})^{\mathrm{T}}\boldsymbol{E}.
\end{align}
Then, similar to (\ref{equations}), if we take $Q$ predetermined DOAs into account, equation (\ref{bsingle11}) can be extended to a group of linear equations as
\begin{align}\label{overall1}
\boldsymbol{B}_{n_1,n_2}\boldsymbol{\gamma}_{n_1,n_2}=\boldsymbol{p}_{n_1,n_2},
\end{align}
where $\boldsymbol{B}_{n_1,n_2}$ is a $Q\times (\frac{2M}{N}-1)$ matrix given by
\begin{align}
\boldsymbol{B}_{n_1,n_2}=\left[\boldsymbol{b}_{n_1,n_2}^{(0)},\boldsymbol{b}_{n_1,n_2}^{(1)},\cdots,\boldsymbol{b}_{n_1,n_2}^{(Q-1)})\right]^{\mathrm{T}}.
\end{align}

Different from (\ref{inverse}) where a diagonal loading coefficient is adopted to avoid the ill-conditioned result, equation (\ref{overall1}) can be solved without diagonal loading if the predetermined DOAs are carefully selected, as will be discussed in the next subsection. Therefore, $\boldsymbol{\gamma}_{n_1,n_2}$ in (\ref{overall1}) can be obtained as
\begin{align}\label{inverse1}
\widehat{\boldsymbol{\gamma}}_{n_1,n_2}=(\boldsymbol{B}_{n_1,n_2}^{\mathrm{H}}\boldsymbol{B}_{n_1,n_2})^{-1}\boldsymbol{B}_{n_1,n_2}^{\mathrm{H}}\boldsymbol{p}_{n_1,n_2}.
\end{align}
Consequently, the sub-SCM can be reconstructed as follows
\begin{align}
\widehat{\boldsymbol{R}}_{n_1,n_2}[m_1,m_2]=\widehat{\gamma}_{n_1,n_2}[m_1-m_2].
\end{align}\par

Similar to (\ref{inverse}), the operator $(\boldsymbol{B}_{n_1,n_2}^{\mathrm{H}}\boldsymbol{B}_{n_1,n_2})^{-1}\boldsymbol{B}_{n_1,n_2}^{\mathrm{H}}$ in (\ref{inverse1}) can be pre-calculated off-line so that the computational burden in (\ref{inverse1}) is mainly caused by the matrix-vector production. Since $\boldsymbol{B}_{n_1,n_2}$ is a $Q\times (\frac{2M}{N}-1)$ matrix, the number of complex multiplications required in (\ref{inverse1}) is only $Q(2M/N-1)$, which is much lower than that required in (\ref{inverse}). In addition to the reduction of complexity, (\ref{overall1}) also indicates that at least $\frac{2M}{N}-1$ predetermined DOAs are required to achieve accurate reconstruction, that is,
$
Q\geq \frac{2M}{N} - 1.
$
This is because there are $\frac{2M}{N}-1$ non-repeated unknowns in $\boldsymbol{\gamma}_{n_1,n_2}$, at least $\frac{2M}{N}-1$ equations are required to solve (\ref{overall1}) with each equation corresponding to one predetermined DOA.
\subsection{Selection of Predetermined DOAs}
In \cite{SLi}, the predetermined DOAs are selected as uniformly distributed from $-90^{\circ}$ to $90^{\circ}$. Although simple, we will show in this subsection that the selection of the predetermined DOAs can be further optimized. Denote
\begin{align}
v^{(q)}=\frac{d}{\lambda}\sin\theta^{(q)}=0.5\sin\theta^{(q)},
\end{align}
to be the space frequency corresponding to $\theta^{(q)}$. In this paper, the predetermined DOAs are selected such that $v^{(q)}$ are uniformly distributed from $-0.5$ to $0.5$, that is, $v^{(q)}=-0.5+q/Q$. As a result, the predetermined DOAs are determined as
\begin{align}\label{theta}
\theta^{(q)}=\arcsin(-1+2q/Q).
\end{align}
In following, it will show $\boldsymbol{B}_{n_1,n_2}^{\mathrm{H}}\boldsymbol{B}_{n_1,n_2}$ can be converted to a diagonal matrix using the predetermined DOAs in (\ref{theta}) so that the matrix inverse operation in (\ref{inverse1}) can be avoided.

If using (\ref{aKro}) and (\ref{matE}) to (\ref{defb}), we have
\begin{align}\label{bsingle}
(\boldsymbol{b}_{n_1,n_2}^{(q)})^{\mathrm{T}}&=(\boldsymbol{a}_{n_2}^{\mathrm{T}}(\theta^{(q)})\otimes\boldsymbol{a}_{n_1}^{\mathrm{H}}(\theta^{(q)}))\cdot\boldsymbol{E}\nonumber\\
&=[a_{0,n_2}(\theta^{(q)})\boldsymbol{a}_{n_1}^{\mathrm{H}}(\theta^{(q)}),\cdots,a_{\frac{M}{N}-1,n_2}(\theta^{(q)})\boldsymbol{a}_{n_1}^{\mathrm{H}}(\theta^{(q)})]\nonumber\\
&\cdot \left[\begin{array}{ccc}
                                               \boldsymbol{O}_{\frac{M}{N} - 1} & \boldsymbol{I}_{\frac{M}{N}} & \boldsymbol{O}_0 \\
                                               \vdots & \vdots & \vdots \\
                                             \boldsymbol{O}_{0} & \boldsymbol{I}_{\frac{M}{N}} & \boldsymbol{O}_{\frac{M}{N} - 1}
                                             \end{array}
\right]\nonumber\\
=&\sum_{m=0}^{\frac{M}{N}-1}a_{m,n_2}(\theta^{(q)})\boldsymbol{a}_{n_1}^{\mathrm{H}}(\theta^{(q)})\left[\boldsymbol{O}_{\frac{M}{N}-1-m},\boldsymbol{I}_{\frac{M}{N}},\boldsymbol{O}_m\right]\nonumber\\
=&\sum_{m=0}^{\frac{M}{N}-1}\left[\boldsymbol{0}_{\frac{M}{N}-1-m}^{\mathrm{T}},a_{m,n_2}(\theta^{(q)})\boldsymbol{a}_{n_1}^{\mathrm{H}}(\theta^{(q)}),\boldsymbol{0}_m^{\mathrm{T}}\right],
\end{align}
where $\boldsymbol{0}_m$ is an $m\times 1$ all-zero vector. If denote ${b}_{n_1,n_2}^{(q)}[m_0]$ as the $m_0$-th entry of $\boldsymbol{b}_{n_1,n_2}^{(q)}$, then we have $m_0 = 0,1,\cdots,\frac{2M}{N} - 2$ since the length of $\boldsymbol{b}_{n_1,n_2}^{(q)}$ is $\frac{2M}{N} - 1$. To proceed, define $\widetilde{a}_{m,n}$ as a sequence with infinite length where
\begin{align}
\widetilde{a}_{m,n}=\begin{cases}
{a}_{m,n}(\theta^{(q)}), & m=0,1,\cdots,\frac{M}{N}-1,\\
0, & \mathrm{others},
\end{cases}.
\end{align}
Then, from the last equation of (\ref{bsingle}), it is easy to verify that ${b}_{n_1,n_2}^{(q)}[m_0]$ can be obtained as
\begin{align}\label{b}
{b}_{n_1,n_2}^{(q)}[m_0]=\sum_{m=0}^{\frac{M}{N}-1}\widetilde{a}_{\frac{M}{N}-1-m_0+m,n_2}\cdot \widetilde{a}_{m,n_1}^*,
\end{align}
which is essentially a linear convolution between $\widetilde{a}_{m_0,n_1}^*$ and $\widetilde{a}_{\frac{M}{N}-1-m_0,n_2}$. Using (\ref{b}), the $(m_1,m_2)$-th entry of $\boldsymbol{B}_{n_1,n_2}^{\mathrm{H}}\boldsymbol{B}_{n_1,n_2}$ with $m_1,m_2=0,1,\cdots,\frac{2M}{N}-2$ is given by
\begin{align}\label{summ}
&[\boldsymbol{B}_{n_1,n_2}^{\mathrm{H}}\boldsymbol{B}_{n_1,n_2}]_{(m_1,m_2)}=\sum_{q=0}^{Q-1}[(\boldsymbol{b}_{n_1,n_2}^{(q)})^*(\boldsymbol{b}_{n_1,n_2}^{(q)})^{\mathrm{T}}]_{(m_1,m_2)}\nonumber\\
&=\sum_{q=0}^{Q-1}b_{n_1,n_2}^{(q)*}[m_1]\cdot b_{n_1,n_2}^{(q)}[m_2]\nonumber\\
&=\sum_{i_1=0}^{\frac{M}{N}-1}\sum_{i_2=0}^{\frac{M}{N}-1}\sum_{q=0}^{Q-1}\widetilde{a}_{\frac{M}{N}-1-m_1+i_1,n_2}^*\widetilde{a}_{i_1,n_1}\widetilde{a}_{\frac{M}{N}-1-m_2+i_2,n_2}\widetilde{a}_{i_2,n_1}^*.
\end{align}
Since the sequence $\widetilde{a}_{m,n}$ has non-zero values only when $m=0,1,\cdots,\frac{M}{N}-1$, $i_1$ in (\ref{summ}) should satisfy
\begin{align}\label{i1}
&0\leq i_1\leq M/N-1,\\
&0\leq M/N - 1-m_1+i_1\leq M/N-1,
\end{align}
simultaneously. Therefore, we can obtain
\begin{align}\label{i1Const}
U_1^{-}\leq i_1\leq U_1^+,
\end{align}
where $U_1^{-}=\mathrm{max}\{0,m_1-(\frac{M}{N}-1)\}$ and $U_1^{+}=\mathrm{min}\{m_1,\frac{M}{N}-1\}$.
Similarly, we have
\begin{align}\label{i2Const}
U_2^{-}\leq i_2\leq U_2^+,
\end{align}
where $U_2^{-}=\mathrm{max}\{0,m_2-(\frac{M}{N}-1)\}$ and $U_2^{+}=\mathrm{min}\{m_2,\frac{M}{N}-1\}$. With the constraints of $i_1$ and $i_2$, (\ref{summ}) can be rewritten as
\begin{align}\label{Bmm}
&[\boldsymbol{B}_{n_1,n_2}^{\mathrm{H}}\boldsymbol{B}_{n_1,n_2}]_{(m_1,m_2)}=\sum_{i_1=U_1^{-}}^{U_1^{+}}\sum_{i_2=U_2^{-}}^{U_2^{+}}\sum_{q=0}^{Q-1}e^{j2\pi v_q(m_1-m_2)},
\end{align}
where we have used the identity
\begin{align}
&a_{\frac{M}{N}-1-m_1+i_1,n_2}^*(\theta^{(q)})a_{i_1,n_1}(\theta^{(q)})\cdot \nonumber\\
&a_{\frac{M}{N}-1-m_2+i_2,n_2}(\theta^{(q)})a_{i_2,n_1}^*(\theta^{(q)})=e^{j2\pi v_q(m_1-m_2)}.
\end{align}
If applying the predetermined DOAs in (\ref{theta}) to (\ref{Bmm}), we have
\begin{align}
\sum_{q=0}^{Q-1}e^{j2\pi (-0.5 + q/Q)(m_1-m_2)}=\begin{cases}
Q, & m_1 = m_2 \\
0, & m_1\neq m_2
\end{cases}.
\end{align}
As a result, we have $[\boldsymbol{B}_{n_1,n_2}^{\mathrm{H}}\boldsymbol{B}_{n_1,n_2}]_{(m_1,m_2)}=0$ for $m_1\neq m_2$ and when $m_1=m_2=0,1,\cdots,2\frac{M}{N}-2$,
\begin{align}
&[\boldsymbol{B}_{n_1,n_2}^{\mathrm{H}}\boldsymbol{B}_{n_1,n_2}]_{(m_1,m_1)}=Q(U_1^+ - U_1^-+1)^2\nonumber\\
&=\begin{cases}
Q(m_1+1)^2, & m_1\leq \frac{M}{N}-1\\
Q\left[2\left(\frac{M}{N} - 1\right) - m_1 + 1\right]^2, & m_1 > \frac{M}{N}-1
\end{cases}.
\end{align}
Apparently, $\boldsymbol{B}_{n_1,n_2}^{\mathrm{H}}\boldsymbol{B}_{n_1,n_2}$ has been converted into a diagonal matrix by selecting predetermined DOAs as in (\ref{theta}), and correspondingly, the matrix inverse operation in (\ref{inverse1}) can be avoided completely.

\section{Simulation Results}
Computer simulation is adopted in this section to investigate the proposed algorithms. We consider a ULA with $M=64$ antennas and the distance between antennas is $0.5\lambda$. $32$ signals are impinging onto the ULA where DOAs of the signals are uniformly distributed from $-90^{\circ}$ to $90^{\circ}$. The arrival signals are assumed independent with zero means and unit powers. Without specification, the signal-to-noise ratio (SNR) is $-5$ dB, and the predetermined DOAs are as in (\ref{theta}). Similar to \cite{SLi}, normalized-square-error (NSE) is used in the simulation to evaluate the accuracy of reconstructed SCM, that is
$
\mathrm{NSE}=\|\widehat{\boldsymbol{R}}-\boldsymbol{R}\|_{\mathrm{F}}^2\cdot \|\boldsymbol{R}\|_{\mathrm{F}}^{-2}.
$
To demonstrate the effectiveness of reconstructed SCM, classical MUSIC algorithm is adopted in the simulation for DOA estimation. Accordingly, means-quare-error (MSE) is used to evaluate the accuracy of DOA estimation, that is
$
\mathrm{MSE}=\mathrm{E}\{|\widehat{\theta}-\theta|^2\}.
$
\begin{figure}
  \centering
  \includegraphics[width=3.5in]{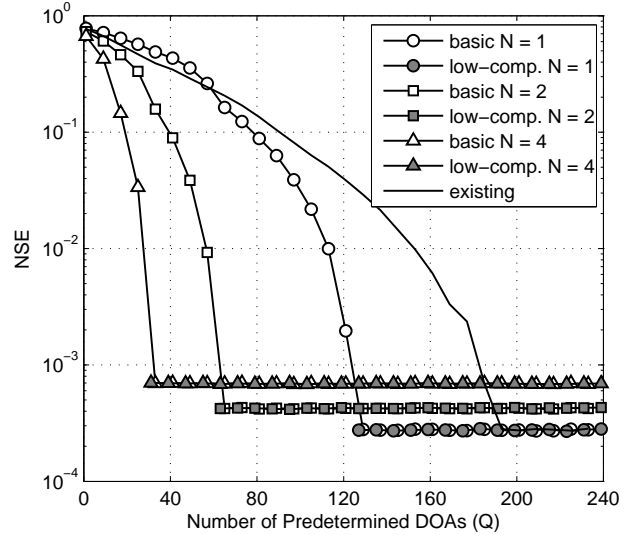}
  \caption{NSE versus the number of predetermined DOAs.}\label{fig1}
\end{figure}
For the basic reconstruction beam sweeping algorithm as in (\ref{inverse}), the diagonal loading coefficient is fixed as $\sigma^2=1$.

\begin{figure*}
  \centering
  \includegraphics[width=2.4in]{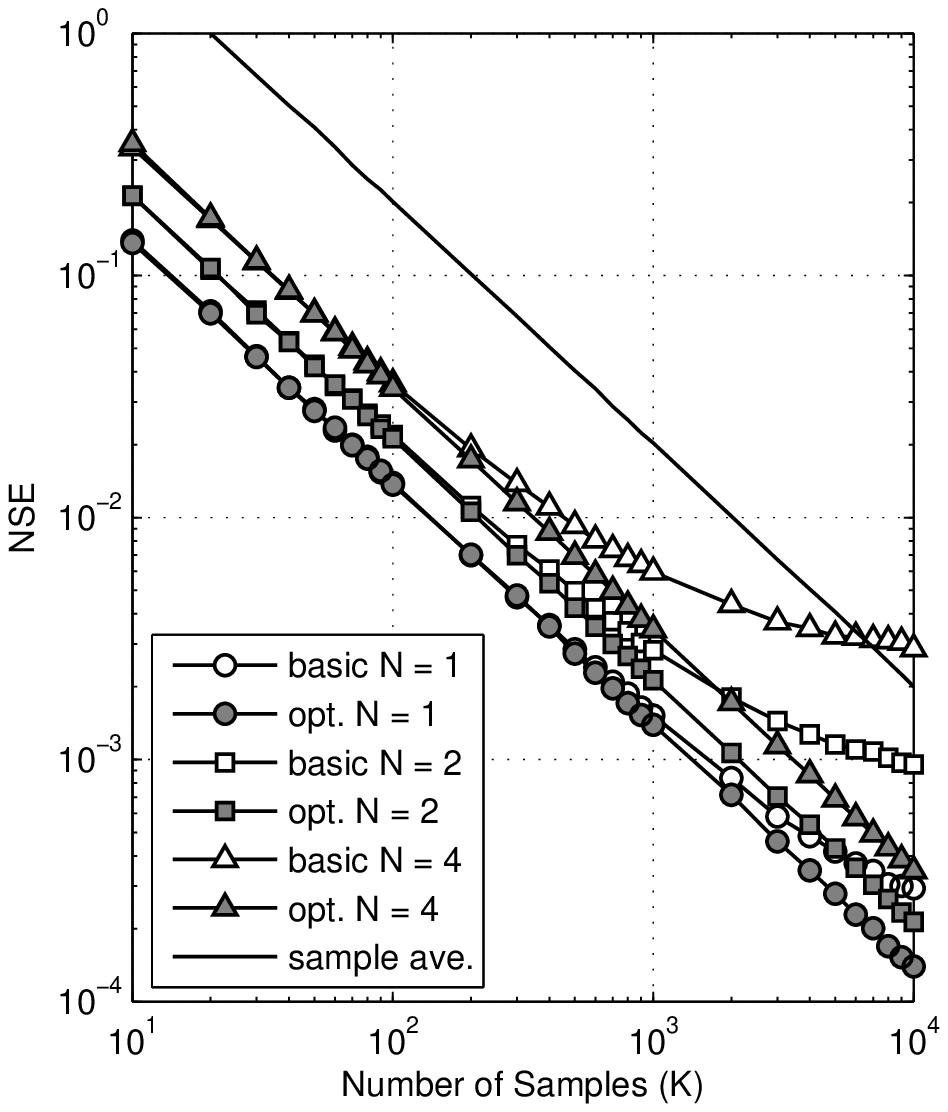}\includegraphics[width=2.4in]{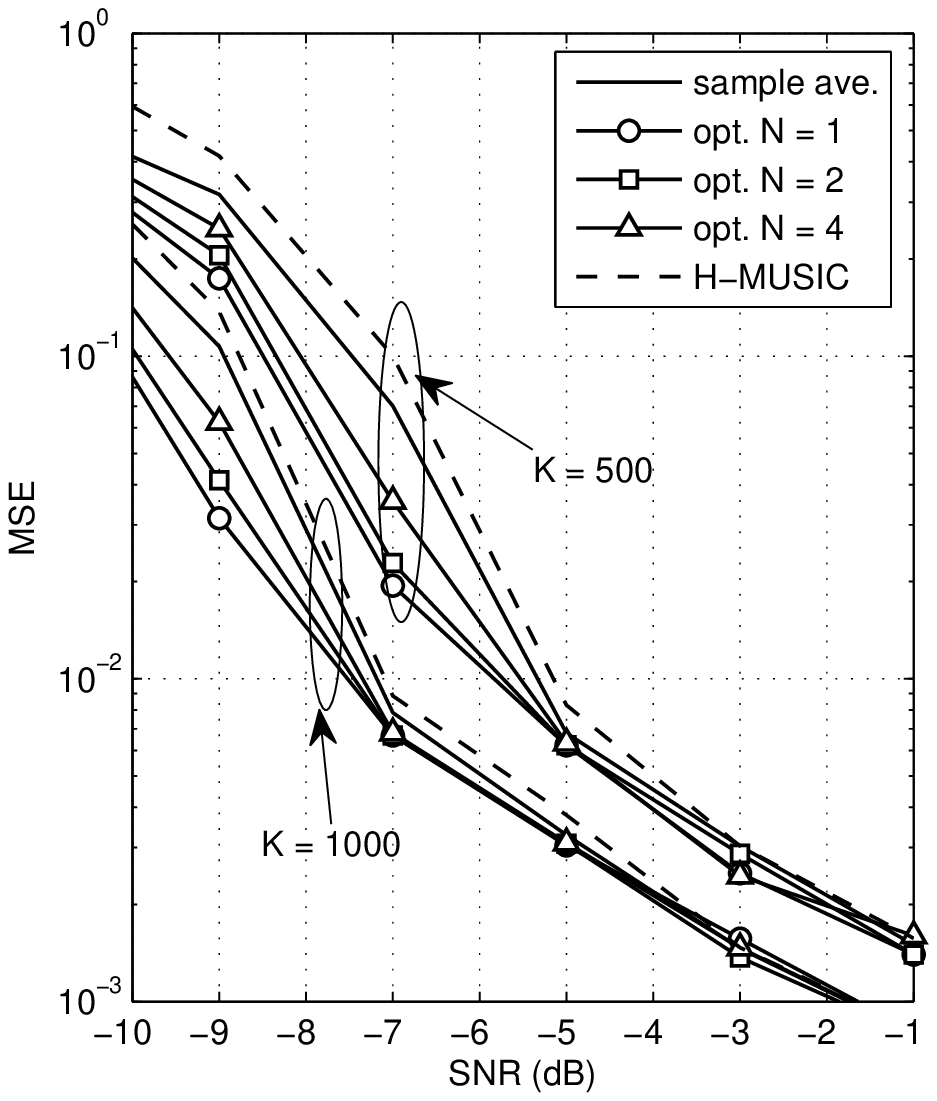}\includegraphics[width=2.4in]{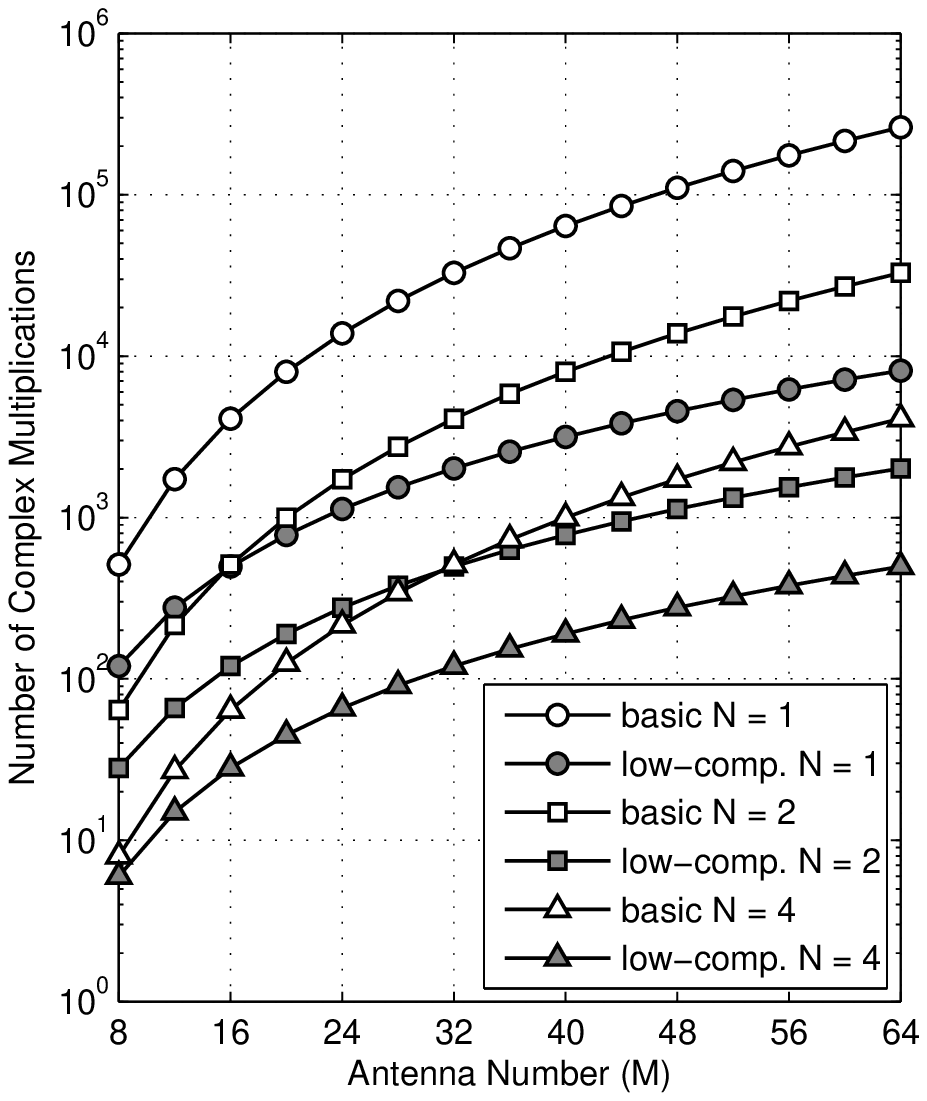}
  (a)~~~~~~~~~~~~~~~~~~~~~~~~~~~~~~~~~~~~~~~~~~~(b)~~~~~~~~~~~~~~~~~~~~~~~~~~~~~~~~~~~~~~~~~~~~(c)
  \caption{(a) NSE versus the sample number (b) MSE of MUSIC using reconstructed SCM (c) number of complex multiplications versus antenna number.}\label{figAll}
\end{figure*}

Fig.~\ref{fig1} shows the reconstruction accuracy with different numbers of predetermined DOAs. The number of samples is fixed as $K=5000$. For the basic algorithm, when $Q$ is small, the reconstruction accuracy can be improved significantly as the rising of $Q$. When $Q>\frac{2M}{N}-1$, there have been sufficient predetermined DOAs and thus the NSE can be hardly reduced further by increasing $Q$. For the low-complexity algorithm as in (\ref{inverse1}), when $Q<\frac{2M}{N}-1$, $\boldsymbol{B}_{n_1,n_2}^{\mathrm{H}}\boldsymbol{B}_{n_1,n_2}$ is ill-conditioned, and thus the low-complexity algorithm is only available when $Q \geq \frac{2M}{N}-1$. It shows that the low-complexity algorithm can achieve minimum NSE when $Q=\frac{2M}{N}-1$. Therefore, the reconstruction procedure can be accomplished within $(\frac{2M}{N}-1)K$ samples. For comparison, Fig.~\ref{fig1} also presents the existing algorithm for the single RF chain case \cite{SLi}. Different from (\ref{theta}), the predetermined DOAs are selected as uniformly distributed from $-90^{\circ}$ to $90^{\circ}$. In this situation, more predetermined DOAs are required to achieve the performances of the proposed algorithms in this paper, leading to a much longer procedure for SCM reconstruction.\par

Fig.~\ref{figAll} (a) shows the reconstruction accuracy with different number of samples. The number of predetermined DOAs is fixed as $Q=\frac{2M}{N}-1$. For the low-complexity algorithm, the NSE can be reduced as the increasing of the number of samples. For the basic algorithm ($N=4$ in particular), however, the reconstruction accuracy can be hardly improved when $K$ is large enough. This observation actually coincides with the result in Fig.~\ref{fig1}. Fig.~\ref{figAll} (a) also shows that better accuracy can be obtained with a smaller number of RF chains. This is because $Q\cdot K$ samples are adopted in overall to reconstruct the SCM. As $Q=\frac{2M}{N}-1$, more samples will be adopted when $N$ is small, and thus the NSE will be improved. For comparison, the traditional sample average algorithm is also included. As expected, the sample average algorithm has the worst performance because it adopts only $K$ samples.\par

Fig.~\ref{figAll} (b) shows the performance of MUSIC algorithm based on the reconstructed SCM. The sample-average based MUSIC algorithm \cite{ROSchmidt2} and the H-MUSIC algorithm \cite{SFChuang} are also included for comparison. For the proposed algorithm, the number of predetermined DOAs is fixed as $Q=\frac{2M}{N}-1$. Since the NSE can be reduced as the reduction of the number of the RF chains, the DOA estimation accuracy can be also improved for massive MIMO systems with small number of RF chains. It also shows that the proposed algorithm can achieve even better performance than sample-average based MUSIC algorithm. This is because the sample average approach is not as accurate as the proposed SCM reconstruction algorithm, as shown in Fig.~\ref{figAll} (a). Although H-MUSIC algorithm can be also used for DOA estimation in hybrid massive MIMO, the estimation accuracy is even worse than the classic MUSIC algorithm. This observation coincides with the result in \cite{SFChuang}.

A comparison on the computation complexity for the basic beam sweeping algorithm and the low-complexity algorithm is shown in Fig.~\ref{figAll} (c). As expected, the low-complexity algorithm in this paper can reduce the computational burden due to the reduction of dimension of matrix-vector production. It also shows that the complexity reduction is more significant when the number of RF chains is smaller.

\section{Conclusions}
In this paper, a beam sweeping approach has been proposed to reconstruct the SCM, so that MUSIC algorithm can be applied for DOA estimation in hybrid massive MIMO systems with multiple RF chains. We have presented the basic algorithm that can be used in the case with multiple RF chains. Then, a low-complexity algorithm has been also introduced by removing the repeated entries in the SCM. In addition, the selection of the predetermined DOAs has also been optimized and the matrix inversion in the low-complexity algorithm can be further avoided. Simulation results have shown that the proposed approach can achieve better performance than existing baselines, and the performance of MUSIC algorithm can be also improved accordingly.

\bibliographystyle{IEEEtran}
\bibliography{IEEEabrv,MUSIC}

\begin{thebibliography}{10}
\providecommand{\url}[1]{#1}
\csname url@samestyle\endcsname
\providecommand{\newblock}{\relax}
\providecommand{\bibinfo}[2]{#2}
\providecommand{\BIBentrySTDinterwordspacing}{\spaceskip=0pt\relax}
\providecommand{\BIBentryALTinterwordstretchfactor}{4}
\providecommand{\BIBentryALTinterwordspacing}{\spaceskip=\fontdimen2\font plus
\BIBentryALTinterwordstretchfactor\fontdimen3\font minus
  \fontdimen4\font\relax}
\providecommand{\BIBforeignlanguage}[2]{{%
\expandafter\ifx\csname l@#1\endcsname\relax
\typeout{** WARNING: IEEEtran.bst: No hyphenation pattern has been}%
\typeout{** loaded for the language `#1'. Using the pattern for}%
\typeout{** the default language instead.}%
\else
\language=\csname l@#1\endcsname
\fi
#2}}
\providecommand{\BIBdecl}{\relax}
\BIBdecl

\bibitem{TETuncer}
T.~E. Tuncer and B.~Friedlander, \emph{Classical and Modern
  Direction-of-Arrival Estimation}.\hskip 1em plus 0.5em minus 0.4em\relax
  Academic, Orlando, 2009.

\bibitem{FShu1}
F.~Shu, X.~Wu, J.~Hu, J.~Li, R.~Chen, and J.~Wang, ``Secure and precise
  wireless transmission for random-subcarrier-selection-based directional
  modulation,'' \emph{IEEE J. Sel. Areas Commun.}, vol.~36, no.~4, pp.
  890--904, July 2018.

\bibitem{ROSchmidt2}
R.~O. Schmidt, ``Multiple emitter location and signal parameter estimation,''
  \emph{IEEE Trans. Antennas Propag.}, no.~3, pp. 276--280, Mar. 1986.

\bibitem{EGLarsson}
E.~G. Larsson, F.~Tufvesson, O.~Edfors, and T.~L. Marzetta, ``Massive {MIMO}
  for next generation wireless systems,'' \emph{IEEE Commun. Mag.}, vol.~52,
  no.~2, pp. 186--195, Feb. 2014.

\bibitem{WRoh}
W.~R. \emph{et al}., ``Millimeter-wave beamforming as an enabling technology
  for {5G} cellular communications: Theoretical feasibility and prototype
  results,'' \emph{IEEE Commun. Mag.}, vol.~52, no.~2, pp. 106--113, Feb. 2014.

\bibitem{LYou}
L.~You, X.~Q. Gao, G.~Y. Li, X.-G. Xia, and N.~Ma, ``{BDMA} for
  millimeter-wave/{T}erahertz massive {MIMO} transmission with per-beam
  synchronization,'' \emph{IEEE J. Sel. Areas Commun.}, vol.~35, no.~7, pp.
  1550--1563, Jul. 2017.

\bibitem{OEAyach1}
O.~E. Ayach, S.~Rajagopal, S.~Abu-Surra, Z.~Pi, and R.~W. Heath, ``Spatially
  sparse precoding in millimeter wave {MIMO} systems,'' \emph{IEEE Trans.
  Wireless Commun.}, vol.~13, no.~3, pp. 1499--1513, Mar. 2014.

\bibitem{VVen}
V.~Venkateswaran and A.~J. van~der Veen, ``Analog beamforming in {MIMO}
  communications with phase shift networks and online channel estimation,''
  \emph{IEEE Trans. Signal Process.}, vol.~58, no.~8, pp. 4131--4143, Aug.
  2010.

\bibitem{CLin}
C.~{Lin} and G.~Y. {Li}, ``Adaptive beamforming with resource allocation for
  distance-aware multi-user indoor {T}erahertz communications,'' \emph{IEEE
  Trans. Commun.}, vol.~63, no.~8, pp. 2985--2995, Aug 2015.

\bibitem{SChuang}
S.~Chuang, W.~Wu, and Y.~Liu, ``High-resolution {AoA} estimation for hybrid
  antenna arrays,'' \emph{IEEE Trans. Antennas Propag.}, vol.~63, no.~7, pp.
  2955--2968, July 2015.

\bibitem{FShu}
F.~Shu, Y.~Qin, T.~Liu, L.~Gui, Y.~Zhang, J.~Li, and Z.~Han, ``Low-complexity
  and high-resolution {DOA} estimation for hybrid analog and digital massive
  {MIMO} receive array,'' \emph{IEEE Trans. Commun.}, vol.~66, no.~6, pp.
  2487--2501, June 2018.

\bibitem{KAgha}
K.~Aghababaiyan, V.~Shah-Mansouri, and B.~Maham, ``High-precision {OMP}-based
  direction of arrival estimation scheme for hybrid non-uniform array,''
  \emph{IEEE Commun. Lett.}, vol.~24, no.~2, Feb. 202.

\bibitem{DHu}
D.~Hu, Y.~Zhang, L.~He, and J.~Wu, ``Low-complexity deep-learning-based {DOA}
  estimation for hybrid massive {MIMO} systems with uniform circular arrays,''
  \emph{IEEE Wirel. Commun. Lett.}, vol.~9, no.~1, pp. 83--86, Jan. 2020.

\bibitem{SLi}
S.~Li, Y.~Liu, L.~You, W.~Wang, H.~Duan, and X.~Li, ``Covariance matrix
  reconstruction for {DOA} estimation in hybrid massive {MIMO} systems,''
  \emph{IEEE Wirel. Commun. Lett.}, vol. Early Acces.

\bibitem{GMan}
D.~G. Manolakis, \emph{Statistical and Adaptie Signal Processing}.\hskip 1em
  plus 0.5em minus 0.4em\relax ARTech House, 2005.

\bibitem{XZhangTsingHua}
X.~Zhang, \emph{Matrix Analysis and Applications (1st Edition)}.\hskip 1em plus
  0.5em minus 0.4em\relax Cambridge University Press, 2017.

\bibitem{RaoBD}
B.~D. Rao and K.~V.~S. Hari, ``Performance analysis of root-{MUSIC},''
  \emph{IEEE Trans. Signal Process.}, vol.~37, no.~12, pp. 1939--1949, 1987.

\bibitem{SFChuang}
S.~F. Chuang, W.~R. Wu, and Y.~T. Liu, ``High-resolution {A}o{A} estimation for
  hybrid antenna arrays,'' \emph{IEEE Trans. Antennas Propag.}, vol.~63, no.~7,
  pp. 2955--2968, July 2015.

\end{thebibliography}

\end{document}